\documentstyle [12pt] {article}

\catcode`@=12
\begin{document}
\thispagestyle{empty}
\begin{flushright} IC/00/65\\UCRHEP-T277\\May 2000\
\end{flushright}
\vspace{0.5in}
\begin{center}
{\Large	\bf Neutrino Masses and the Gluino Axion Model\\}
\vspace{1.5in}
{\bf D. A. Demir$^1$, Ernest Ma$^2$, and Utpal Sarkar$^{2,3}$\\}
\vspace{0.5in}
{$^1$ \sl The Abdus Salam International Centre for Theoretical Physics, 
Trieste 34100, Italy\\}
\vspace{0.1in}
{$^2$ \sl Department of Physics, University of California, Riverside, 
California 92521, USA\\}
\vspace{0.1in}
{$^3$ \sl Physical Research Laboratory, Ahmedabad 380 009, India\\}
\vspace{1.5in}
\end{center}
\begin{abstract}\
We extend the recently proposed gluino axion model to include neutrino 
masses.  We discuss how the canonical seesaw model and the Higgs triplet 
model may be realized in this framework.  In the former case, the heavy 
singlet neutrinos are contained in superfields which do not have any vacuum 
expectation value, whereas the gluino axion is contained in one which does. 
We also construct a specific renormalizable model which realizes the mass 
scale relationship $M_{SUSY} \sim f_a^2/M_U$, where $f_a$ is the axion 
decay constant and $M_U$ is a large effective mass parameter.
\end{abstract}

\newpage
\baselineskip 24pt

A new axionic solution\cite{pqww} to the strong CP problem was recently 
proposed\cite{gluax}.  Instead of coupling to ordinary matter as in the 
DFSZ model\cite{dfsz} or to unknown matter as in the KSVZ model\cite{ksvz}, 
this new axion couples to the gluino as well as all other supersymmetric 
particles.  The instanton-induced CP violating phase\cite{theta} of quantum 
chromodynamics is then canceled by the dynamical phase of the gluino mass, 
as opposed to that of the quarks in the DFSZ model and that of the unknown 
colored fermions in the KSVZ model.  This means that CP violation is absent 
in the strong-interaction sector and experimental observables, such as the 
neutron electric dipole moment\cite{nedm}, are subject only to 
weak-interaction contributions.

What sets the gluino axion model\cite{gluax} apart from all other previous 
models is its identification of the Peccei-Quinn global symmetry $U(1)_{PQ}$ 
with the $U(1)_R$ symmetry of superfield transformations.  Under $U(1)_R$, 
the scalar components of a chiral superfield transform as $\phi \to e^{i 
\theta R} \phi$, whereas the fermionic components transform as $\psi \to 
e^{i \theta (R-1)} \psi$.  In the Minimal Supersymmetric Standard Model 
(MSSM), the quark and lepton superfields $\hat Q$, $\hat u^c$, $\hat d^c$, 
$\hat L$, $\hat e^c$ have $R = +1$ whereas the Higgs superfields $\hat H_u$, 
$\hat H_d$ have $R = 0$.  The superpotential
\begin{equation}
\hat W = \mu \hat H_u \hat H_d + h_u \hat H_u \hat Q \hat u^c + h_d \hat H_d 
\hat Q \hat d^c + h_e \hat H_d \hat L \hat e^c
\end{equation}
has $R = +2$ except for the $\mu$ term (which has $R = 0$).  Hence the 
resulting Lagrangian breaks $U(1)_R$ explicitly, leaving only a discrete 
remnant, i.e. the usual $R$ parity: $R = (-1)^{3B+L+2J}$.  The gluino 
axion model replaces $\mu$ with a singlet composite superfield of $R = +2$ 
so that the resulting supersymmetric Lagrangian is invariant under $U(1)_R$. 
It also requires all supersymmetry breaking terms to be invariant under 
$U(1)_R$, the spontaneous breaking of which then produces the axion and 
solves the strong CP problem.

In the MSSM, neutrinos are massless.  However, in view of the recent 
experimental evidence for neutrino oscillations, it is desirable to 
incorporate into any realistic model naturally small Majorana neutrino 
masses\cite{wein,path}.  In the following we will discuss how the canonical 
seesaw model\cite{seesaw} and the Higgs triplet model\cite{triplet} may be 
realized in the framework of the gluino axion model.  In the case of the 
seesaw model, there are in fact proposals\cite{nuax} that the axion scale 
is the same as that of the singlet neutrino masses.

Consider first the Higgs triplet model.  Add to the gluino axion model 
two triplet superfields:
\begin{eqnarray}
\hat \xi_1 &=& (\xi_1^{++}, \xi_1^+, \xi_1^0): ~~~ R = 0, \\ 
\hat \xi_2 &=& (\xi_2^0, \xi_2^-, \xi_2^{--}): ~~~ R = +2,
\end{eqnarray}
then the superpotential (which is required to have $R = +2$) has the 
following additional terms:
\begin{equation}
\Delta \hat W = m_\xi \hat \xi_1 \hat \xi_2 + f_{ij} \hat \xi_1 \hat L_i 
\hat L_j + h \hat \xi_2 \hat H_u \hat H_u.
\end{equation}
Note that the term $\hat \xi_1 \hat H_d \hat H_d$ is forbidden.  The 
resulting scalar potential has the term $|m_\xi \xi_1 + h H_u H_u|^2$, 
hence the desired trilinear scalar interaction $h m_\xi \xi_1^\dagger 
H_u H_u + h.c.$ is there to combine with the Yukawa interaction $f_{ij} 
\xi_1 L_i L_j + h.c.$ to form the well-known dimension-5 effective 
operator\cite{wein} which generates the neutrino masses:
\begin{equation}
(m_\nu)_{ij} = 2 f_{ij} h {\langle H_u \rangle^2 \over m_\xi}.
\end{equation}
If the intermediate scale $m_\xi$ is assumed to be of order the $U(1)_R$ 
breaking scale, i.e. $10^{11}$ GeV or so, then $m_\nu$ of order 1 eV is 
obtained if $f_{ij} h$ is of order $10^{-2}$.

Consider next the canonical sesaw model.  Add to the gluino axion model 
the singlet superfield $\hat N$ with $R=+1$, then the superpotential is 
supplemented by 
\begin{equation} 
\Delta \hat W = m_N \hat N \hat N + f_{i} \hat L_i \hat N \hat H_u~,
\end{equation}
which generates the well-known seesaw neutrino mass 
\begin{equation}
(m_\nu)_{ij} = f_{i}f_{j} {\langle H_u \rangle^2 \over m_N}~.
\end{equation}

Since both $\hat N$ and $\hat S$ have the same $U(1)_R$ charge, it is 
tempting to identify them as one, so that its scalar component has a large 
vacuum expectation value (VEV) and contains the axion, while its fermionic 
component is the heavy neutrino singlet of mass $m_N$.  However, the 
resulting scalar potential will now contain the term $|2 m_N \tilde N + f_i 
\tilde L_i H_u|^2$, so that the scalar bilinear term $\tilde L_i H_u$ (which 
violates lepton number) has the huge coefficient $2 f_i m_N \langle \tilde N 
\rangle$ which is clearly unacceptable.  To prevent $\hat N$ from picking 
up any VEV, we introduce the discrete symmetry $L$ parity, under which 
$\hat L$, $\hat e^c$, and $\hat N$ are odd and all other superfields are 
even, including $\hat S$.

In proposing the gluino axion model\cite{gluax}, the composite operator 
$\mu (\hat S) \equiv (\hat S)^2/M_{Pl}$ with $R = +2$ is used.  The couplings 
of $\mu (S)$ to the supersymmetric particles of the MSSM are required to be 
invariant under $U(1)_R$.  Hence the supersymmetry of the MSSM is broken 
by $\mu_{eff} = \langle S \rangle^2/M_{Pl}$.  In the following we consider 
an alternative scheme, using the fundamental singlet superfields $\hat S_2$, 
$\hat S_1$, and $\hat S_0$, with $R = 2,1,0$ respectively.  We impose the 
discrete symmetry $Z_3$ with $\omega^3 = 1$ on all superfields as follows:
\begin{eqnarray}
1: && \hat u^c, \hat d^c, \hat e^c, \hat N, \\ 
\omega: && \hat Q, \hat L, \hat S_1, \hat S_0, \hat \xi_1\\ 
\omega^2: && \hat H_u, \hat H_d, \hat S_2, \hat \xi_2.
\end{eqnarray}
We see then that Eqs.~(4) and (6) are allowed in addition to Eq. (1) except 
for the $\mu$ term.  The superpotential involving $\hat S_2$, $\hat S_1$, 
and $\hat S_0$ is required to have $R = +2$ also:
\begin{equation}
\hat W = m_2 \hat S_2 \hat S_0 + f \hat S_1 \hat S_1 \hat S_0 + h \hat S_2 
\hat H_u \hat H_d.
\end{equation}
The resulting scalar potential is
\begin{equation}
V = |m_2 S_0 + h H_u H_d|^2 + |2 f S_1 S_0|^2 + |m_2 S_2 + f S_1 S_1|^2.
\end{equation}
Let $v_i \equiv \langle S_i \rangle$, then $V = 0$ has the solution
\begin{equation}
v_0 = 0, ~~~ v_2 = {-f v_1^2 \over m_2}.
\end{equation}
The problem now is of course the indeterminate value\cite{inde} of $v_1$. 
To fix $v_1$ and maintain the above seesaw structure while keeping $v_0$ 
zero, we add the following soft terms:
\begin{eqnarray}
V' =  -{m'_1}^2 |S_1|^2 - [\lambda m_2 S_2^* S_1^2 + h.c].
\end{eqnarray}
The equations of constraint for $V + V'$ to be a minimum are
\begin{eqnarray}
0 &=& m_2^2 v_2 + (f-\lambda) m_2 v_1^2, \\ 0 &=& -{m'_1}^2 + 4 f^2 v_0^2 + 
2(f-\lambda) m_2 v_2 + 2 f^2 v_1^2, \\ 0 &=& v_0 (m_2^2 + 4 f^2 v_1^2). 
\end{eqnarray}
From Eq.~(15), we find
\begin{equation}
v_2 = {(\lambda - f) v_1^2 \over m_2},
\end{equation}
which indeed preserves the expected seesaw structure.  From Eq.~(17), we 
see that $v_0=0$ is still a solution, and from Eq.~(16), taking into account 
Eq.~(18), we find
\begin{equation}
v_1^2 = {{m'_1}^2 \over 2 \lambda (2 f - \lambda)},
\end{equation}
where the denominator must be positive for $V + V'$ to be a minimum.  The 
discrete $Z_3$ symmetry is broken spontaneously by $v_1$, hence a possible 
domain wall problem may appear.  However, the Majorana fermion singlet 
$\tilde S_1$ may be given a mass $m_1$ which breaks the $Z_3$ symmetry 
softly but explicitly, thus avoiding such a problem.

Note that the scalars $S_0$ and $S_2$ remain heavy with mass $m_2$, but 
their VEV's are zero or very small\cite{triplet,lima}.  The global $U(1)_R$ 
symmetry is broken by $\langle S_1 \rangle$ and $\langle S_2 \rangle$, hence 
the resulting Nambu-Goldstone boson\cite{nago} is given by
\begin{equation}
{(v_1) \sqrt 2 Im S_1 + (2 v_2) \sqrt 2 Im S_2 \over \sqrt {v_1^2 + 4 v_2^2}}.
\end{equation}
In the couplings of $S_2$ to the superparticles of the MSSM, the axion enters 
as $S_2$ is replaced by
\begin{equation}
v_2 e^{2i \varphi} = v_2 e^{2i \langle \varphi \rangle} 
\exp \left( {ia \sqrt 2 \over v}
\right),
\end{equation}
where $v = \sqrt {v_1^2 + 4 v_2^2}$ and the axion $a$ is given by
\begin{equation}
a = (\sqrt 2 v) [\varphi - \langle \varphi \rangle],
\end{equation}
with $\langle \varphi \rangle = - \theta_{QCD}/6$.  Thus the axion decay 
constant $f_a$ is $\sqrt 2 v \simeq \sqrt 2 v_1$ but $M_{SUSY}$ of the MSSM 
is $v_2$.  This is analogous to the DFSZ model\cite{dfsz} with $M_{SUSY}$ 
replaced by $M_W$.

In this model, the seesaw condition of Eq.~(18) implies that $M_{SUSY} \sim 
f_a^2/M_U$, where $M_U$ is a large effective mass parameter, i.e. $2m_2/h
(\lambda - f)$.  The allowed range of values for $f_a$ from 
astrophysics and cosmology\cite{rafe} is between $10^9$ and $10^{12}$ GeV.  
Hence $M_U$ is between $10^{15}$ and $10^{21}$ GeV.  Neutrino masses are 
given by either Eq.~(5) in the Higgs triplet model, or Eq.~(7) in the 
canonical seesaw model. There is no {\it a priori} connection between $f_a$ 
and $m_\xi$ or $m_N$. However, if they are of the same order of magnitude, 
then $m_\nu$ is inversely proportional to $f_a$ as proposed in the models 
of Ref.\cite{nuax}.

The laboratory detection\cite{robi} of axions depends on the $a \to \gamma 
\gamma$ coupling, which is proportional to\cite{rafe}
\begin{equation}
{E \over N} - \left( {2 \over 3} \right) {4 + m_u/m_d + m_u/m_s \over 1 + 
m_u/m_d + m_u/m_s} = {E \over N} - 1.92 \pm 0.08,
\end{equation}
where $N$ and $E$ are coefficients proportional to the color and 
electromagnetic anomalies of the axion.  For the gluino axion model, $N=6$ 
but $E=0$ without or with neutrino mass from either the canonical seesaw or 
the Higgs triplet mechanism.  This comes from the fact that, except for the 
gluino, every left-handed fermion has a right-handed partner of the 
same $R$.

In conclusion, we have incorporated neutrino masses (through the canonical 
seesaw or Higgs triplet mechanism) into the gluino axion model, using 
the superpotentials of Eq.~(1) [without the $\mu$ term] and Eq.~(11) with 
either Eq.~(6) or Eq.~(4).  The $\mu$ term is replaced by $h \hat S_2$, 
so that
\begin{equation}
\mu_{eff} = h \langle S_2 \rangle = h v_2 e^{-i\theta_{QCD}/3}.
\end{equation}
Assuming that the intermediate scales, i.e. $v_1$ and $m_N$ or $m_\xi$ are 
of the same order, we then have
\begin{equation}
M_{SUSY} \sim |\mu_{eff}| \sim {f_a^2 \over M_U}, ~~~ m_\nu \sim f_{eff}^2 
{M_W^2 \over f_a},
\end{equation}
where $f_{eff}$ is a dimensionless coupling.

~

\begin{center} {ACKNOWLEDGEMENT}
\end{center}

We thank A. Yu. Smirnov for reading the manuscript.  The research of E.M. 
was supported in part by the U.~S.~Department of Energy under Grant 
No.~DE-FG03-94ER40837.

\bibliographystyle{unsrt}

\end{document}